\documentclass{article}
\usepackage{ijcai24}

\usepackage{times}
\usepackage{soul}
\usepackage{url}
\usepackage[hidelinks]{hyperref}
\usepackage[utf8]{inputenc}
\usepackage[small]{caption}
\usepackage{algorithmic}

\usepackage[ruled,vlined,lined,ruled,linesnumbered]{algorithm2e}
\usepackage{graphicx}
\usepackage{amsmath}
\usepackage{amsthm}
\newtheorem{definition}{Definition}
\usepackage{booktabs}

\SetKwInput{KwInput}{Input}                
\SetKwInput{KwOutput}{Output}              

\newlength\mylen 

\title{Privacy-preserving Quantification of Non-IID Degree in Federated Learning}

\author{
Yuping Yan$^{1,3}$
\and
Yizhi Wang$^1$\and
Yingchao Yu$^{2}$\And
Yaochu Jin$^{1,*}$\\
\affiliations
$^1$School of Engineering, Westlake University\\
$^2$College of Information Science and Technology, Donghua University\\
$^3$Faculty of Informatics, Department of Computeralgebra, Eötvös Loránd University\\
$^*$ Corresponding author\\
\emails
\{yanyuping, wangyizhi, jinyaochu\}@westlake.edu.cn,
yingchaoyuu@outlook.com
}

\begin{document}

\maketitle

\begin{abstract}
Federated learning (FL) offers a privacy-preserving approach to machine learning for multiple collaborators without sharing raw data. However, the existence of non-independent and non-identically distributed (non-IID) datasets across different clients presents a significant challenge to FL, leading to a sharp drop in accuracy, reduced efficiency, and hindered implementation. To address the non-IID problem, various methods have been proposed, including clustering and personalized FL frameworks. Nevertheless, to date, a formal quantitative definition of the non-IID degree between different clients' datasets is still missing, hindering the clients from comparing and obtaining an overview of their data distributions with other clients. For the first time, this paper proposes a quantitative definition of the non-IID degree in the federated environment by employing the cumulative distribution function (CDF), called Fully Homomorphic Encryption-based Federated Cumulative Distribution Function (FHE-FCDF). This method utilizes cryptographic primitive fully homomorphic encryption to enable clients to estimate the non-IID degree while ensuring privacy preservation. The experiments conducted on the CIFAR-100 non-IID dataset validate the effectiveness of our proposed method.
\end{abstract}

\section{Introduction}

Federated Learning (FL) \cite{konevcny2016federated} is a decentralized machine learning paradigm that allows multiple clients (e.g., devices, edge nodes, or users) to collaboratively train a shared machine learning model while keeping their data locally stored and private. The central server orchestrates the training process by aggregating the model updates from individual clients, effectively incorporating knowledge from diverse data sources while preserving data privacy and security. This enables the creation of robust and generalized models that can adapt to different client-specific data distributions and characteristics. 

Despite its potential, FL faces numerous unsolved challenges, with statistical heterogeneity (i.e., non-IID data)~\cite{zhu2021federated,ma2022state} and privacy~\cite{lyu2020threats,yin2021comprehensive} emerging as critical issues hindering its development. In machine learning, non-IID data refers to scenarios where data distributions are independent and identically distributed from each other. In such environments, the FL framework may experience a loss of accuracy and struggle to achieve model convergence. Experiments by Chai et al. \cite{Chai2020FedEvalAB} demonstrate that FedAvg, one of the classic FL algorithms, can experience a significant decline in accuracy of up to 9\% when dealing with non-IID datasets. This finding underscores the adverse impact of non-IID data on model accuracy, leading to issues such as instability and poor generalization ability. Researchers have proposed various schemes and models to address these shortcomings, with personalizing local models~\cite{arivazhagan2019federated,zhang2021parameterized} and employing more robust algorithms~\cite{karimireddy2020scaffold,gao2022feddc} for learning an effective global model emerging as the two most effective and popular approaches. However, little research has been published that quantifies the non-IID degree of different datasets before applying the FL process.

While traditional data distribution metrics like mean, variance, skewness, or kurtosis provide insights into the level of non-IIDness in datasets, these methods often involve direct data analysis. Statistics approaches such as Kullback–Leibler divergence \cite{van2014renyi}, Jensen-Shannon divergence \cite{menendez1997jensen}, and cross-entropy \cite{de2005tutorial} offer means to measure differences between probability distributions. However, these analyses may necessitate data sharing or centralized processing, presenting substantial risks to data privacy and security.

The FL framework offers a promising solution to address this issue while circumventing the need for local data sharing. However, the current landscape provides limited comprehensive solutions to tackle the intricate challenge of quantifying non-IIDness within the FL framework. The study conducted by Li et al. \cite{li2019convergence} introduces a convergence-based technique, which involves incorporating the minimum value of the global objective function with the minimum value of individual local objective functions. In \cite{zhao2018federated}, the authors define the weight divergence, which is the standard way to quantify the non-IID degree of different datasets. However, except for these specific approaches, a generic method for defining the non-IIDness of datasets distributed to different clients is missing. Even within the FL framework, where local data remains non-public and is not directly shared, security and privacy could still be compromised due to the exchange of model parameters. This exchange of information introduces potential vulnerabilities and attacks, such as gradients leakage \cite{zhu2019deep} and interference attacks \cite{shokri2017membership}. As a result, quantifying the non-IIDness of different datasets in a privacy-preserving federated manner becomes even more challenging.

To ensure privacy preservation in FL, additional security building blocks are required, including secure aggregation and multi-party computation protocols, as well as cryptographic primitives like Homomorphic Encryption (HE) ~\cite{hao2019efficient,zhang2021dubhe} and Differential Privacy (DP)~\cite{wei2021user,cheng2022differentially}. Implementing these privacy-enhancing techniques becomes imperative to safeguard sensitive information and foster the success of FL. Among these techniques, HE stands out as a notable approach often referred to as the 'Swiss army knife of cryptography.' It enables computations to be conducted on encrypted data without requiring access to the secret decryption key. The results of these computations remain encrypted and can only be revealed by the owner of the secret key. Through the adoption of HE, FL participants can securely share their encrypted model updates without exposing their raw data. The central server can then aggregate these encrypted updates to generate a global model, all while preserving data privacy and confidentiality.

To address the challenge of precisely and comprehensively quantifying the non-IIDness of different datasets in a federated and privacy-preserving manner, we introduce the Fully Homomorphic Encryption-based Federated Cumulative Distribution Function (FHE-FCDF) approach. The Cumulative Distribution Function (CDF) \cite{chun2000uncertainty} represents the probability distribution of a discrete random variable or the cumulative probabilities of the data. By comparing the CDFs of different datasets, we can identify variations in the data distribution among clients. Meanwhile, from the shape and trend of the CDF curves of different clients, we can assess data heterogeneity. The main contributions of this work are:

\begin{itemize}
\item In this framework, we introduce the Cumulative Distribution Function as a precise tool for quantifying the degree of non-IID characteristics across various datasets from different clients.

\item We implement the homomorphic encryption cryptographic scheme in the calculation of Cumulative Distribution Function and federated aggregation, without exposing the raw data of clients.

\item We validate the effectiveness of our proposed quantification method within a federated framework using the non-IID distributed CIFAR-100 dataset, considering both label distribution skew and feature distribution skew.

\end{itemize}

The rest of the paper is organized as follows. Section II presents the preliminaries of this work. The proposed Fully Homomorphic Encryption-based Federated Cumulative Distribution Function (FHE-FCDF) approach is detailed in Section III, followed by the validation in Section IV. Finally, Section V concludes the paper and proposes future work.

\section{Preliminaries}

This section presents the fundamental concepts underlying our proposed method. It encompasses an exploration of federated learning and the drift effect within the context of non-IID conditions, the Cumulative Distribution Function, and the Fully Homomorphic Encryption Scheme.

\subsection{Federated learning and Non-IID}
\begin{figure}[htb]
    \centering
    \includegraphics[width=3.5in]{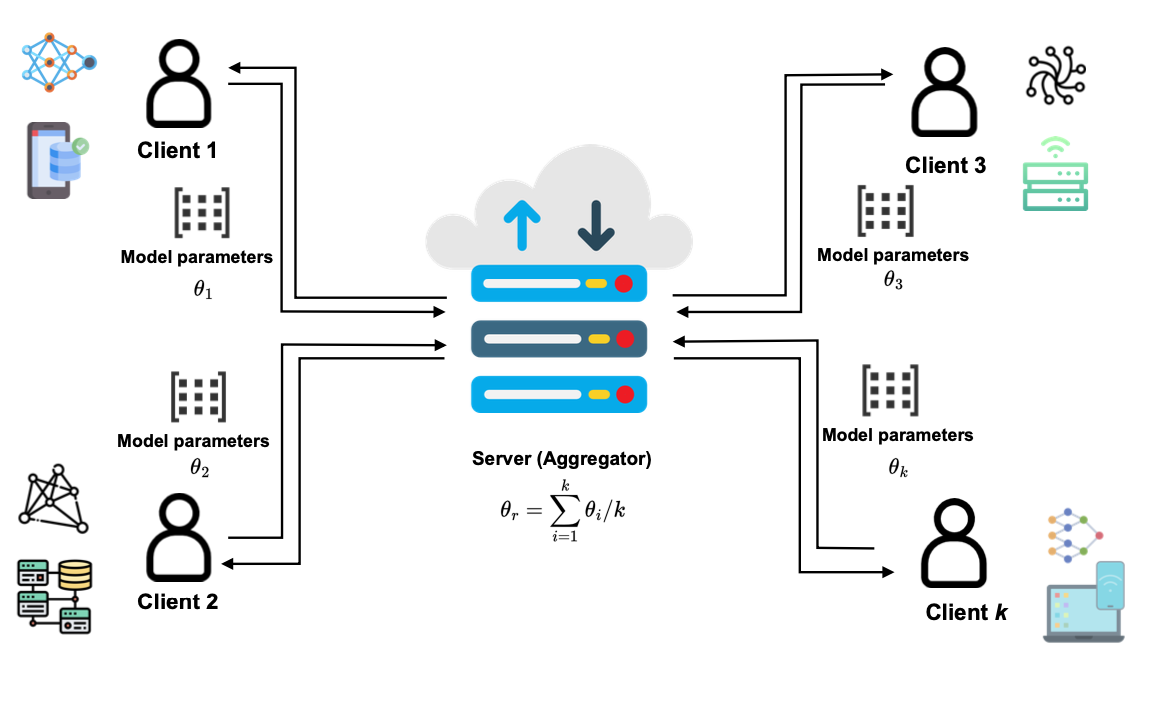}
    \caption{The overview of federated learning framework.}
    \label{fig:fl}
\end{figure}

The classic FL framework operates in a client-server setting, where a central server is responsible for coordinating the training process, and the individual client devices perform model training locally. The illustration of the FL training process can be found in Fig. \ref{fig:fl}.

The basic framework of FL can be summarized as follows:
\begin{definition}[Federated Learning]
    Suppose there are $n$ participants $U_1,U_2,\cdots, U_n$, and each participant $U_i$ has its own local dataset $D_i$. The task is to compute the global model $M_{global}$ on the central server (or aggregator) based on all uploaded local models, while each of the local models is generated from the local client dataset. In the first round, the server will generate a model with random parameters $\theta_0$ and send them to all clients. After receiving the model sent by the server, $k$ out of $n$ participating clients will locally compute training gradients based on their dataset and send the updated model to the server. Then, the server aggregates gradients of clients and computes the global parameters $\theta_r=\sum_{i=1}^{n}\theta_i/n$. After a round of updates is completed, the clients check whether the accuracy of the local model meets the requirements and stop training if it does; otherwise, it is ready for the next round of training.   
\end{definition}   
\subsubsection{Effect of Non-IID Data}
\begin{figure}[htb]
    \centering
    \includegraphics[width=3.8in]{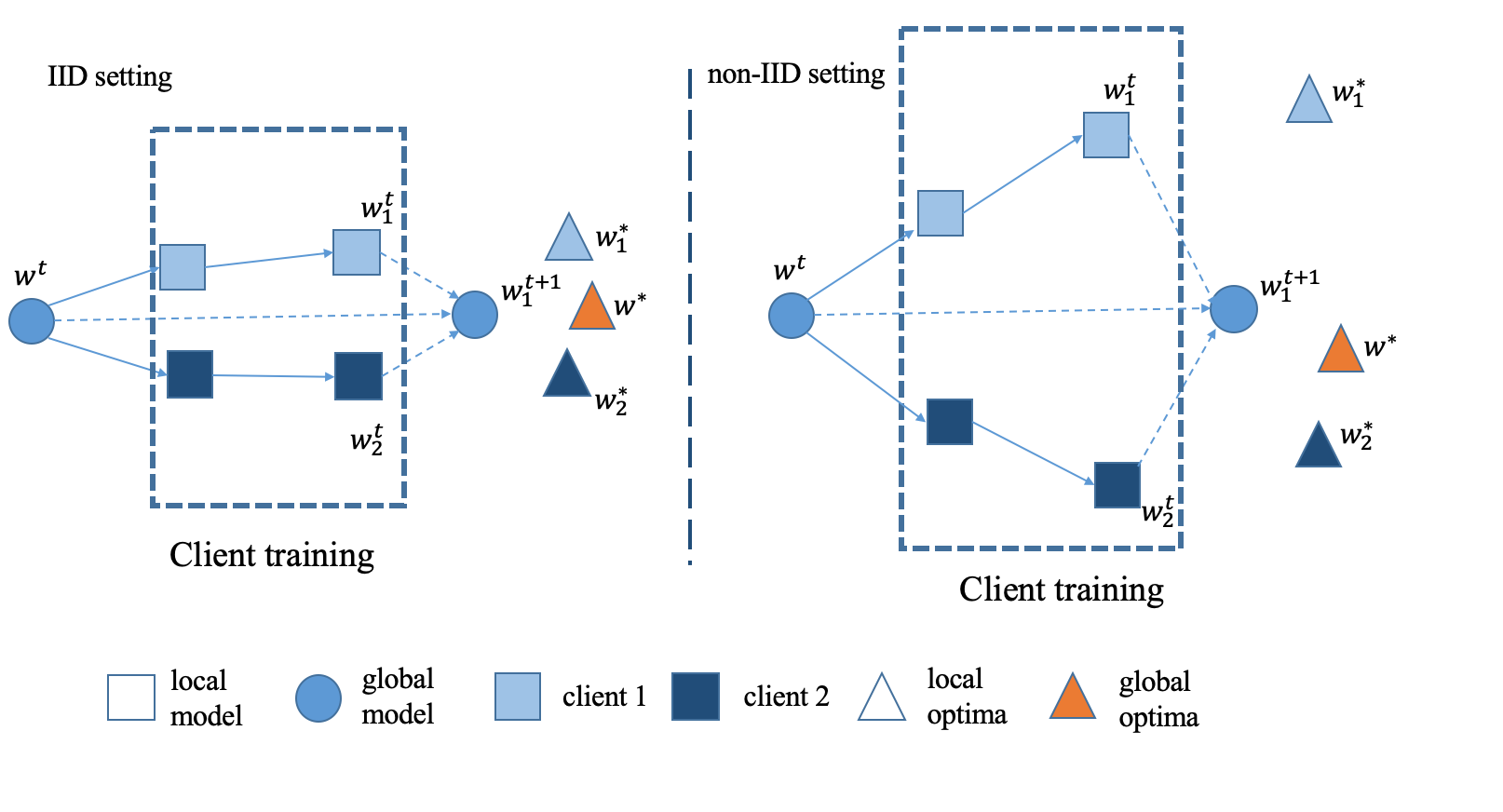}
    \caption{Weight divergence of FL with IID and non-IID data.}
    \label{fig:non-iid}
\end{figure}

As the distribution of the local dataset is different from the global distribution in the non-IID scenario, the local optima will be far from the global optima. Eventually, the global accuracy will drop in non-IID data settings. As shown in Fig. \ref{fig:non-iid}, when the dataset is IID, the divergence between local optima and global optima is usually minor. However, in the non-IID case, $w_1^t$ is far different from $w^*$ due to the distance between the data distribution, making the divergence between $w^*$ and $w_1^*$ much larger.

According to the extensive overview of non-IID data presented by Kairouz et al. \cite{kairouz2021advances}, there exist five distinct typical categories of non-IIDness. These include label distribution skew, feature distribution skew, similar labels with differing features, identical features with diverse labels, and quantity skew. Defining the local data distribution $P(x_i,y_i) = P(y_i|x_i)P(x_i)$ or $P(x_i,y_i) = P(x_i|y_i)P(y_i)$ ($x_i$ and $y_i$ are two different datasets). The research by  \cite{zhu2021federated} also investigates non-IID data on temporal skew.

In this paper, we only consider the label distribution skew and feature distribution skew of non-IID cases. 
\begin{itemize}
    \item \textbf{Label distribution skew:} The label distributions $P(y_i)$ vary across different parties, as each participant may have a distinct set of labels. This unbalanced label distribution can exacerbate model divergence, wherein the model encounters difficulties in effectively generalizing across disparate label patterns.
    \item \textbf{Feature distribution skew:} The feature distributions $P(x_i)$ vary across different parties, as each participant may possess a distinct set of features. These variations in feature distributions pose a challenge in achieving global models that can effectively handle diverse feature representations.
\end{itemize}

\subsection{Empirical Cumulative Distribution Function}

In mathematical terms, the Cumulative Distribution Function (CDF) is a probability function that allows us to find the cumulative probability for a given value. By using the CDF, we can obtain a table describing the probability distribution of a random variable.

The cumulative distribution function of a real-valued random variable $X$ is the function given by \cite{park2018fundamentals}:
\begin{align}
    F_X(x)=\mathrm{P}(X \leq x),
\end{align}
where the right-hand side represents the probability that the random variable $X$ takes on a value less than or equal to $x$. 

The probability of $X$ in this range of a semi-closed interval $(a, b ]$, where $a<b$, can be presented as:
\begin{equation}
\mathrm{P}(a<X \leq b)=F_X(b)-F_X(a)
\end{equation}

An empirical CDF is an estimate of the CDF. The empirical CDF plot is similar to a probability plot except both axes are linear, which can make the empirical CDF plot more intuitive to interpret. 

The empirical CDF is to take each sample point $X_i$ with the probability of $1/n$. The empirical CDF $\hat{F}_n(x)$ is defined as the follows: 
\begin{equation}
    \hat{F}_n(x)=\frac{1}{n} \sum_{i=1}^n I\left(X_i \leq x\right) \text {, while } I\left(X_i \leq x\right)= \begin{cases}1 & X_i \leq x \\ 0 & X_i>x\end{cases}
\end{equation}

An illustration of CDF and empirical CDF can be found in Fig. \ref{fig:ecdf}:

\begin{figure}[htb]
    \centering
    \includegraphics[width=2.5in]{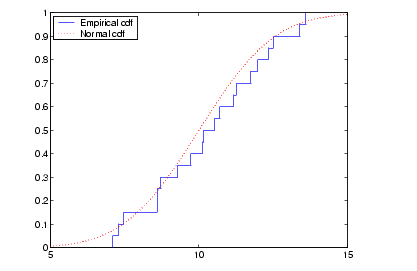}
    \caption{CDF vs. empirical CDF.}
    \label{fig:ecdf}
\end{figure}

\subsection{Fully Homomorphic Encryption Scheme}
Homomorphic encryption is a powerful tool for safeguarding data privacy while simultaneously performing computational tasks. The definition of HE can be found as follows. 
\begin{definition}[Homomorphic Encryption]
Given two sets $A$ and $B$, and a function $f: A \rightarrow B$, $f$ is said to be an additive homomorphism if, for any $x$ and $y$ in $A$, we have: $f(x \diamond y) = f(x) \diamond f(y)$. If the operation $\diamond$ represents an addition, we define it as an additive homomorphism. On the other hand, if $\diamond$ represents multiplication, we define it as a multiplicative homomorphism.
\end{definition}
A fully homomorphic encryption scheme consists of a set of probabilistic polynomial time algorithms defined as follows: $E$ = (KeyGen, Encrypt, Decrypt, Evaluate):
\begin{itemize}
    \item $KeyGen\left(1^\lambda\right) \rightarrow(sk, pk, evk)$: Given the security parameter $\lambda$, outputs a key pair consisting of a public encryption key $pk$, a secret decryption key $sk$, and an evaluation key $evk$.
    \item $Enc(pk, m) \rightarrow ct$: Given a message $m \in \mathcal{M}$ and an encryption key $pk$, outputs a ciphertext $ct$.
    \item $Dec(sk, ct)=m$: Given the secret decryption key and a ciphertext $ct$ encrypting $m$, outputs $m$.
    \item  $Eval\left(evk, f, ct_1, ct_2, \ldots, ct_n\right) \rightarrow ct^{\prime}$: Given the evaluation key, a description of a function $f: \mathcal{M}^n \rightarrow \mathcal{M}$, and $n$ ciphertexts encrypting messages $m_1, \ldots, m_n$, outputs the result ciphertext $ct^{\prime}$ encrypting $m^{\prime}=$ $f\left(m_1, \ldots, m_n\right)$.
\end{itemize}

\begin{figure*}[htb]
    \centering
    \includegraphics[width=7in]{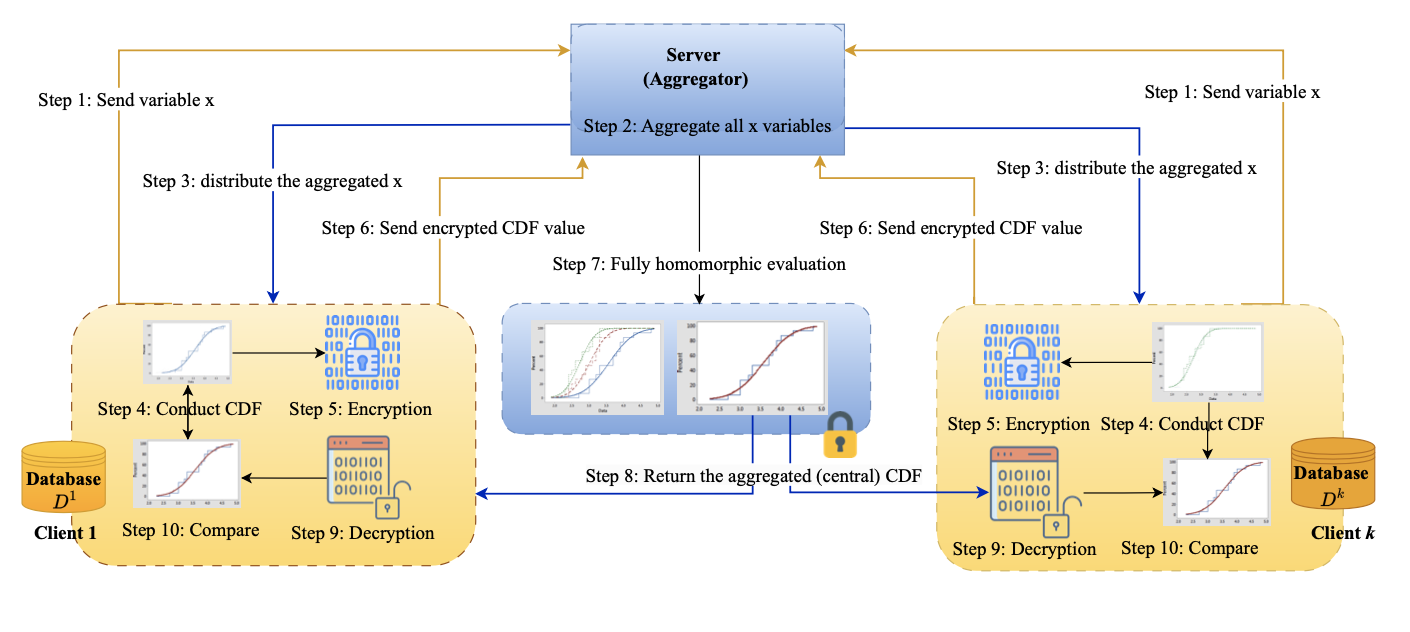}
    \caption{An overview of the FHE-FCDF approach consisting of one server and multiple clients. \textcircled{1}: All clients sent the variable $x$ to the server. \textcircled{2}: The server aggregates all $x$ variables. \textcircled{3}: The server sends the distribution policy of the variable $x$ to all the clients.\textcircled{4}: Clients conduct CDF based on the distribution policy and their local database. \textcircled{5}: Encrypt the CDF values with the encryption key of the fully homomorphic encryption scheme. \textcircled{6}: All the clients send the encrypted CDF values to the server in the ciphertext. \textcircled{7}: The server conducts the fully homomorphic evaluation based on these values and gets the aggregation CDF (central CDF). \textcircled{8}: The server returns the aggregated CDF to all the clients. \textcircled{9}: Each client decrypts the aggregated CDF with the decryption key of the fully homomorphic encryption scheme. \textcircled{10}: The client compares the central CDF with the local CDF and quantifies the non-IID degree.}
    \label{fig:overview}
\end{figure*}

\section{Overview of FHE-FCDF approach}
In this section, we describe the proposed FHE-FCDF algorithm to securely quantify the non-IIDness of different parties with a fully homomorphic encryption scheme and federated framework.

The overview of the FHE-FCDF approach is depicted in Fig. \ref{fig:overview}. The framework follows a server-client structure, and the detailed process is as follows:
\begin{enumerate}
    \item Each client possesses its local database, which may vary in label distributions, feature distributions, and other attributes. Leveraging their local data, clients transmit their variable $x$ to the server. The nature of $x$ can be either labels or features based on their respective local databases.
    \item Since clients may have diverse types of $x$, the server aggregates these values and returns the updated $x$ as the distribution policy.
    \item Upon receiving the aggregated $x$, each client calculates their CDF values based on their raw data. To preserve privacy, the results are encrypted using the encryption (public) key of the FHE scheme.
    \item The server collects the encrypted CDFs from all clients and conducts fully homomorphic evaluations to obtain the average CDF values in an encrypted format. The server then shares the aggregated (central) CDF with all clients.
    \item All clients receive the aggregated CDF and decrypt it using the decryption (private) key of the FHE scheme. By comparing their locally generated CDF with the central CDF, clients can assess the degree of non-IIDness of their local database relative to others.
\end{enumerate}

This approach comprises two main stages: the local non-IID illustration stage and the federated aggregation with fully homomorphic encryption stage. We will provide detailed explanations for both stages in the following sections and summarize them in Algorithm \ref{HE-FCDE alg}.

\begin{figure*}[htb]
    \centering
    \includegraphics[width=7in]{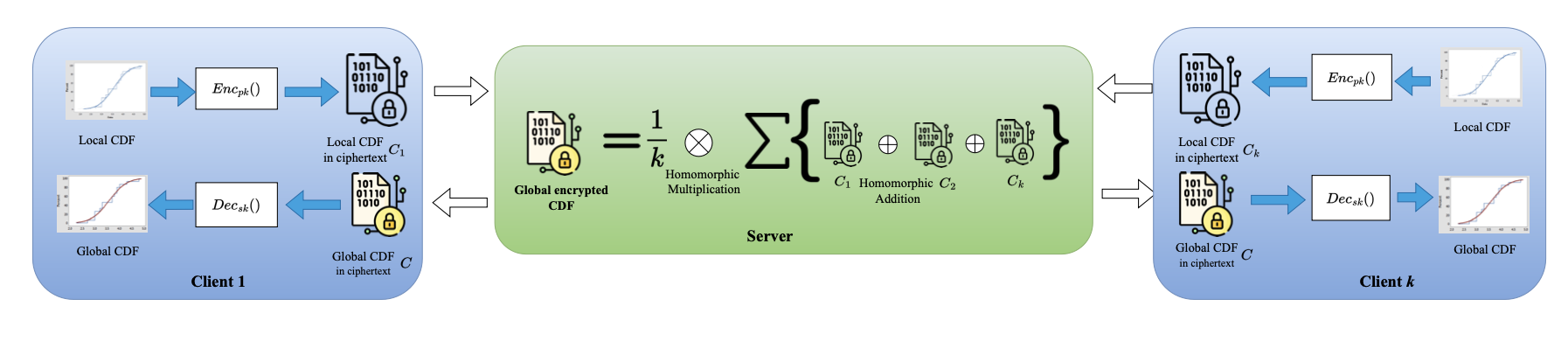}
    \caption{The fully homomorphic evaluation process}
    \label{fig:fhe}
\end{figure*}
\subsection{Local non-IID illustration stage}

To mathematically quantify the non-IIDness of a local dataset, we will employ both the CDF and the empirical CDF (eCDF). The results derived from the CDF will be encrypted and transmitted to the server for homomorphic evaluation. This approach leverages the vector format of the CDF results, which supports FHE vector-based calculations and evaluation. Conversely, the eCDF will enable users to independently and easily compare non-IIDness due to its straightforward expression and illustrative clarity.

For one client, assume we have the input of labels $x_1, x_2 ... x_c$ are independent, identically distributed real random variables, we generate the CDF plot by the following equation:
\begin{equation}
    F(x)=\sum_{c=1}^r \operatorname{Pr}\left(X=x_c\right),
\end{equation}
where $F(x)$ is the definition of CDF as $F(x)=\operatorname{Pr}(X \leq x)$. $x_r$ is the largest possible value of $X$ that is less than or equal to $x$.

To calculate the eCDF, we take each sample point $x_1, x_2 ... x_c$ with the probability of $1/n$. The empirical CDF $\hat{F}_n(x)$ is defined as the follows:
\begin{equation}
 \hat{F}_n(x)=\frac{1}{n} \sum_{c=1}^n I\left(X \leq x_c\right) \text {, while } I\left(X \leq x_c\right)= \begin{cases}1 & X \leq x_c \\ 0 & X>x_c\end{cases}   
\end{equation}

\begin{algorithm}
\normalsize
 \SetAlgoLined
 \SetKwData{Left}{left}\SetKwData{This}{this}\SetKwData{Up}{up}
 \SetKwRepeat{doWhile}{do}{while}
 \SetKwFunction{Union}{Union}\SetKwFunction{FindCompress}{FindCompress}
 \SetKwInOut{Input}{Input}\SetKwInOut{Output}{Output}
 \Input{Security parameter $\lambda$, number of clients $k$, ring size $n$, the polynomial ring $R=Z[X] /\left(X^{n+1}\right)$, a ciphertext modulus $q$, a special modulus $p$ coprime to $q$, a key distribution $\chi$, and an error distribution $\Omega$ over $R$, input value of of client $k$ $\Vec{x^k} = (x_1^k,...x_c^k)$, $c$ is the number of labels, Public key $PK$, Secrete key $sk$}
  \textbf{Key generation:} \\ 
  Sample $\mathbf{s} \leftarrow \chi$ \\
  Sample $\vec{e}  \leftarrow \Omega$\\
  Sample $\mathbf{a} \leftarrow R_q$\\
  Compute the secret key: $sk=\mathbf{s}$ \\
  Compute the public key: $PK=\left([-(\mathbf{a} \cdot \mathbf{s}+\mathbf{e})]_q, \mathbf{a}\right)$ \\
  \textbf{Local client encryption:}\\
   Sample $a \leftarrow R_q$\\
 \For{$\Vec{x} \in R_q^2$}{
  Compute the first part of ciphertext: $\vec{ct[0]^k} = -a$ \\
  Compute the second part of ciphertext: $\vec{ct[1]^k} = a \cdot \mathbf{s}+ \frac{q}{p} \Vec{x_c^k} + \Vec{e}$ \\}
  Combine the encrypted vector of client $k$: $\vec{ct_{x^k}} = [\Vec{ct_0^k},\Vec{ct_1^k}]$ \\
  \textbf{Global server aggregation:} \\
  Compute $\vec{ct_{s}} = \frac{1}{k} \cdot \sum \vec{ct_{x}^k}$ \\
  $\vec{ct[0]_{s}} =  \sum \vec{ct[0]{x}_c^k}$\\
  $\vec{ct[1]_{s}} =  \sum \vec{ct[1]{x}_c^k}$ \\
  \textbf{Local client decryption:} \\
   Compute $\Vec{x_s} = \frac{p}{q}\left(\Vec{ct[0]_s} \cdot \mathbf{s}+ct[1]_s\right)=\Vec{x_s} +\frac{p}{q} e_0$ \\
\caption{FCDF scheme.}
\label{HE-FCDE alg}
\end{algorithm}

\subsection{Fully Homomorphic Evaluation}

To implement the FHE scheme, the procedure comprises four pivotal steps: key generation, local client encryption, global server aggregation, and local client decryption. A comprehensive breakdown of these steps can be found in Algorithm \ref{HE-FCDE alg}, while an overarching representation of the complete FHE evaluation process is depicted in Figure \ref{fig:fhe}.

In the entire process, a third trusted party is utilized to generate the homomorphic encryption secret key $sk$ and public key $PK$ based on the FHE key generation function. All clients are provided with these key pairs for conducting homomorphic evaluations. Each client generates its local CDF $\Vec{x_c}$ based on their local dataset distribution, presented as vectors, and encrypts the vector values using the homomorphic public key $PK$ with the local client encryption function. The server receives all the local CDF values in the ciphertext and conducts aggregation and mean functions to obtain the global CDF vectors. The server then returns the global CDF to all clients in an encrypted format. Thus, in this process, the server cannot learn any information about clients' local CDF values or the local database. Upon receiving the global CDF, all clients decrypt it using the secret key $sk$. Using the global CDF, clients can convert it to the eCDF line, facilitating comparison of the non-IID degree between the local eCDF and the global eCDF.

\section{Empirical Validation}
To validate the effectiveness of our proposed method, we conducted experiments within a federated framework using the non-IID distributed CIFAR-100 dataset ~\cite{krizhevsky2009learning}.

\begin{figure*}[htb]
    \centering
    \includegraphics[width=5in]{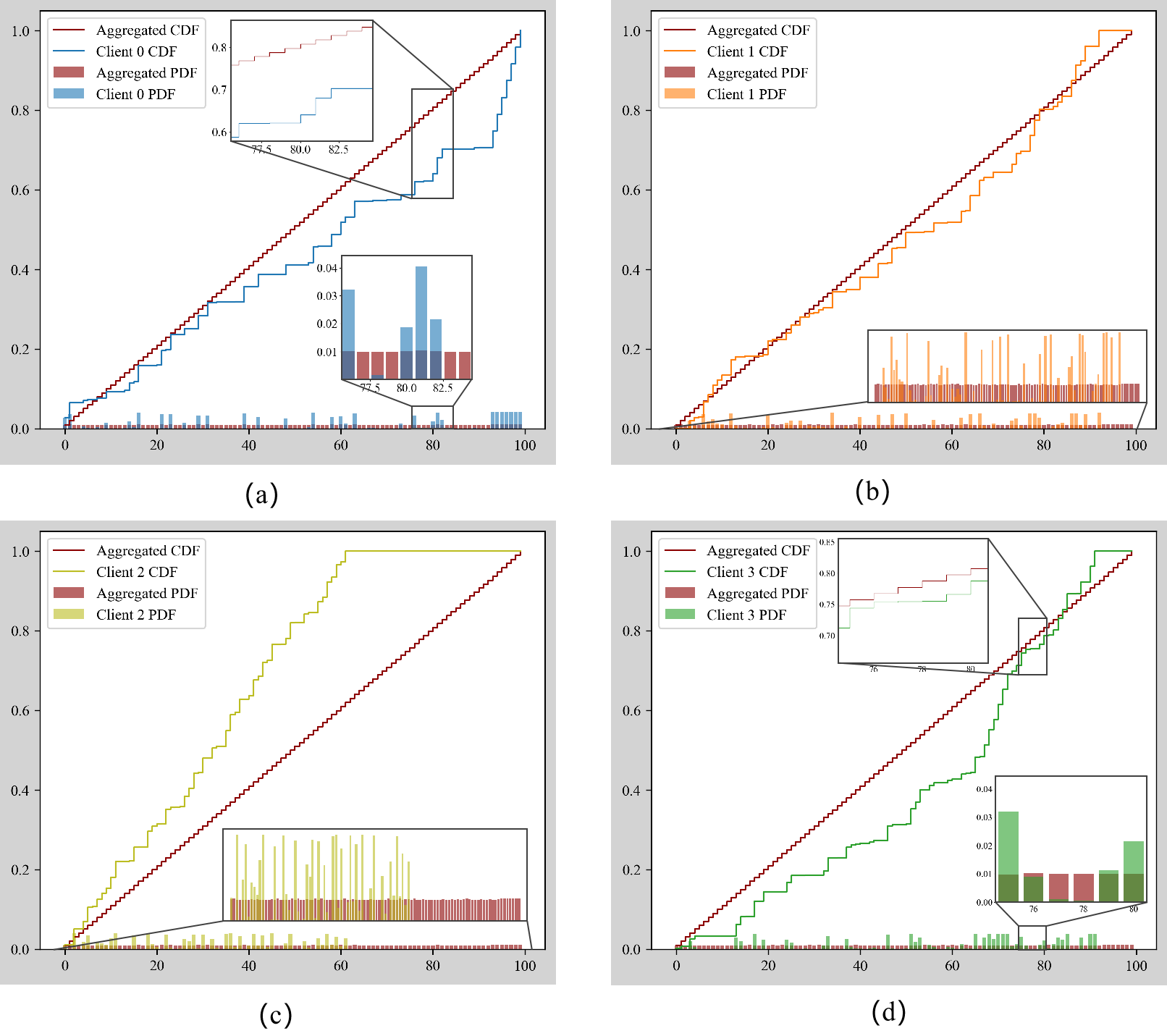}
    \caption{Non-IIDness of labels comparison with aggregated PDF on each client.}
    \label{fig:labels}
\end{figure*}

\subsubsection{Quantification of Label Non-IID}
In assessing label distribution, we set the number of clients $k = 4$, and the local database $D^{i}$ of client $i$ was sampled according to $p(D_j^i) \sim  Dir(0.1)$. The $Dir(\beta)$ denotes the Dirichlet distribution~\cite{yurochkin2019bayesian}, in which a smaller $\beta$ exacerbates the heavier imbalance degree of the distribution, and $p(D_j^i)$ denotes the proportion of sample labelled with $j$ in the local database of client $i$. Here, we set the $\beta$ to $0.1$ to simulate the possible skewness in the practical FL scenarios.

The visualization of comparison on label distribution obtained by FHE-FCDF is shown in Fig.~\ref{fig:labels}. For an intuitive understanding, the probability density functions (PDF) stored locally and derived from the aggregated CDF are also illustrated in the figure, presenting the probability of each label in the database. For example, in Figs.~\ref{fig:labels} (a) and (d), in terms of label proportion, client 1 has a greater proportion on the label “squirrel” (Acronyms as 80), and client 4 exhibits a lower weight on the label “snail” (Acronyms as 77). Further, the vertical lines on CDFs indicate contributions from a client for that particular label, while flatter lines suggest a scarcity of that data. For instance, in Fig.~\ref{fig:labels} (b), despite slight differences in label portions, the relatively closer CDFs reflect a uniform local label distribution, indicating that client 1 may receive a better model benefit from FL. Conversely, Client 2 in Fig.~\ref{fig:labels}  (c) might appear to have uniformly distributed data if observed locally. Unfortunately, based on receiving the global CDF, it shows the client owns only about 60$\%$ of the data labels, indicating collaborative training is possible to provide some outcomes that are not useful for this client.

\begin{figure}[htb]
    \centering
    \includegraphics[width=3in]{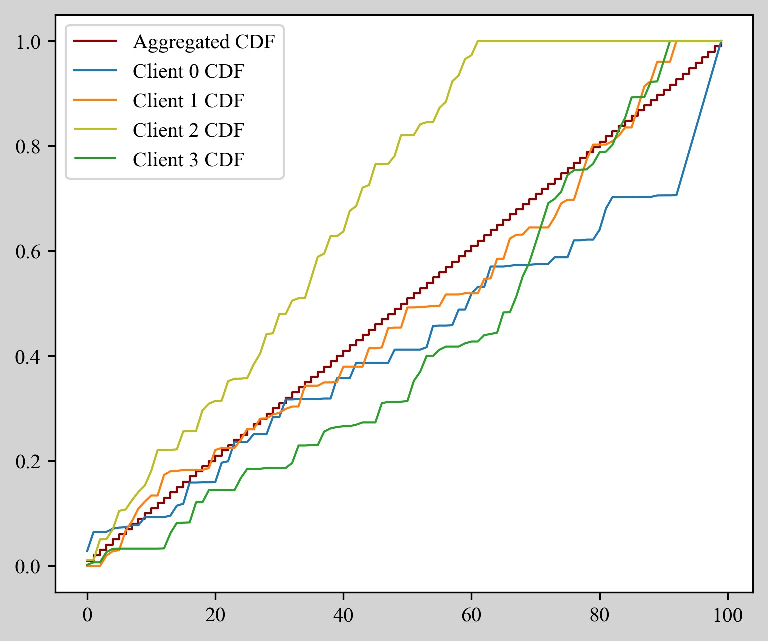}
    \caption{Aggregated CDF and the local CDFs.}
    \label{fig:cdfs}
\end{figure}

To illustrate the differences in local label distributions, we displayed both the aggregated CDF and the local CDFs in Fig.~\ref{fig:cdfs}. However, it is noteworthy that the aggregation is executed within the Full Homomorphic Encryption, which means that the server does not actually have access to real CDFs in practice.

\subsubsection{Quantification of Non-IIDness}
\begin{figure*}[htb]
    \centering
    \includegraphics[width=7in]{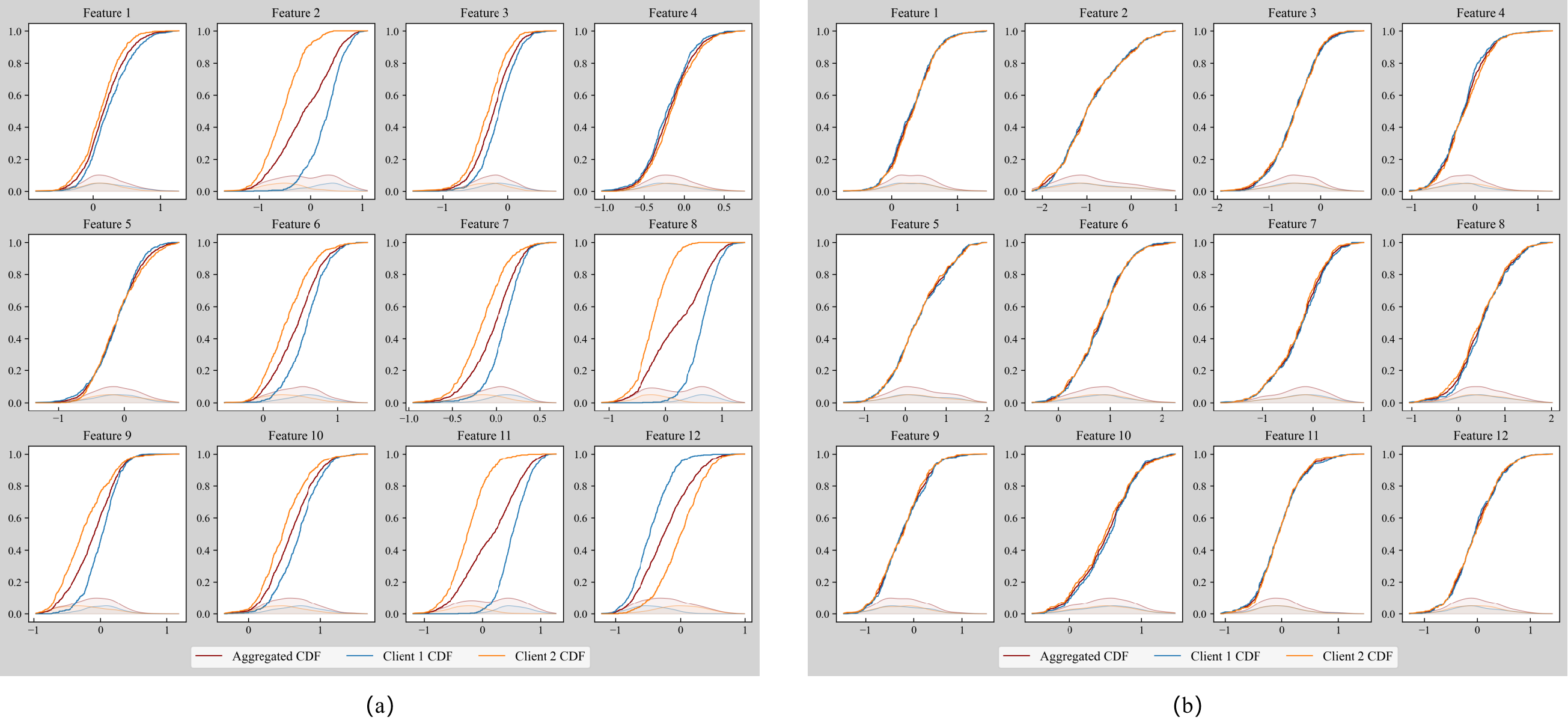}
    \caption{The extracted features' CDFs on (a) Non-IID distributed clients and (b) IID distributed clients comparing with aggregated CDF.}
    \label{fig:features}
\end{figure*}
We employed a pretrained ResNet18~\cite{he2016deep} without the classification layer as the feature extractor. For better visualization, a linear layer was added to reduce the feature vector into 12 dimensions. To preserve the consistency of the semantics of the extracted representations, each dimension was quantified separately to ensure their order remained unchanged. We set $k = 2$. For each client $i$, the size of the local database was set to 300. To simulate the Non-IID scenarios, at least $75\%$ of the data on each client was drawn from distinct fine-grain classes in CIFAR-100. CDFs of features extracted from clients compared with the aggregated one are shown in Fig.~\ref{fig:features} (a). For the IID case in Fig ~\ref{fig:features} (b), at least $75\%$ of the client's data was drawn from the same class without conjunction, while the remaining $20\%$ data was randomly sampled from other classes. Compared with Fig.~\ref{fig:features} (b), discrepancies can be observed on CDFs in most feature dimensions in Fig.~\ref{fig:features} (a). This highlights that it can be captured by the proposed FHE-FCDF if distinct differences exist between data feature distributions.

\section{Conclusion}
In this paper, we propose a metric, for the first time to our best knowledge, for quantifying the non-IID degree in federated learning environments by leveraging FHE. Our approach empowers clients to assess their non-IID degree before initiating the learning process, thereby incentivizing their active participation in the federated update of the models. By providing clients with insights into the non-IID characteristics of their data, our method promotes transparency and fosters collaboration in federated learning settings. The experimental validation of our proposed method on the CIFAR-100 dataset demonstrates its effectiveness in accurately quantifying the non-IID degree across diverse client datasets. In the future, we will adopt the proposed non-IID measure to assist our clients in selecting the most appropriate federated learning frameworks tailored to their specific non-IID characteristics.





\bibliographystyle{named}
\bibliography{ijcai24}

\end{document}